\documentclass[12pt]{article}
\usepackage{graphicx}
\renewcommand{\vec}[1]{{\mathbf{#1}}}
\newcommand{\beq}{\begin{eqnarray}} 
\newcommand{\eeq}{\end{eqnarray}}

\usepackage{scicite}

\usepackage{times}

\newenvironment{sciabstract}{%
\begin{quote} \bf}
{\end{quote}}

\newcounter{lastnote}

\title{The Elusive Bose Metal}

\author
{Philip Phillips,$^{1\ast}$ Denis Dalidovich,$^{2}$\\
\\
\normalsize{$^{1}$Loomis Laboratory of Physics, University of Illinois at Urbana-Champaign,}\\
\normalsize{1100 W.Green St., Urbana, IL., 61801-3080}\\
\normalsize{$^{2}$National High Field Magnetic Laboratory,Florida State University}\\
\normalsize{Tallahassee, Florida 32310}
\\
\normalsize{$^\ast$To whom correspondence should be addressed; E-mail: dimer@uiuc.edu.}
}

\date{}

\begin{document} 


\baselineskip24pt


\maketitle


\begin{sciabstract}
 The conventional theory of metals is in crisis.  In the last 15 years, there 
has been an 
unexpected sprouting of metallic states in low dimensional systems directly contradicting conventional wisdom.  For example, bosons are thought to 
exist in one of two ground states:  condensed in a superconductor or localized in
an insulator.  However, several experiments on thin metal alloy films have 
observed that a metallic phase disrupts the direct transition between
the superconductor and the insulator.  We analyze the experiments on the insulator-superconductor transition and argue that the intervening metallic phase is bosonic.  All relevant theoretical proposals for the Bose metal are discussed, particularly the recent idea that the metallic phase is glassy.  The implications for the putative vortex glass state in the copper-oxide superconductors are examined.
\end{sciabstract}


Before the advent of quantum mechanics, the microscopic origin of electrical conduction in metals remained shrouded in a central mystery: How do electrons avoid the atoms in a dense material?  Realising that electrons move as waves, Bloch\cite{bloch} proposed that electrons surf the atoms in a crystal by
adjusting their wavelength to fit the periodicity of the lattice.  Bloch's view is radical in that electron-electron interactions are absent and defects are assumed to provide only a minor correction.  We know now that both of these assumptions are wrong.  For example, defects can destroy perfect conduction leading to electron localization\cite{go4} and electron interactions can generate new many-body states such as superconductivity and magnetism.  Even in the extreme case in which single electrons\cite{ml,kk} are localized by defect scattering, superconductivity still obtains.  How is this possible?  At low temperatures, lattice-mediated attractive
interactions between electrons produce a resistanceless state in which the
charge carriers are electron pairs, Cooper pairs. Pair formation, however,
is not a sufficient condition for superconductivity.
Superconductivity obtains when all of the Cooper pairs phase lock into a single quantum state.  Such macroscopic occupation of a single quantum state is not possible for fermions as a result of the Pauli exclusion principle. 
However, Cooper pairs whose radius of gyration is smaller than the inter-pair spacing are bosons and there is no exclusion principle for
bosons.  Hence, macroscopic occupation of a single quantum state is 
permissible and it is the resultant phase coherence that
thwarts localization.   Superconductivity in the localized
electron regime can be thought of as Bose-Einstein condensation, a phenomenon which has received much attention recently with the myriad of experiments reporting Bose superfluidity in optical lattices of alkali atoms\cite{weiman,optical}.  In the other extreme where the inter-pair spacing is smaller than the Cooper pair radius, the Cooper pair$\rightarrow$boson mapping breaks down.  Nonetheless, phase coherence still obtains.  Consequently, we will adopt the simplification
in this article that Cooper pairs are simply charge $2e$ bosons, $e$ the electron charge.  

Because bosons can superconduct, one might presuppose that they can also exist in a metallic state.  However, as we will see,
a simple quantum mechanical principle precludes this state of affairs.
In low dimensions ($D\le 2$), except in the presence of correlated disorder\cite{rdm} or a magnetic  field\cite{wl,qhall}, not even non-interacting electrons can remain metallic at zero temperature
in the presence of disorder.  When the interactions are weak and repulsive,
the ground state appears to be ferromagnetic\cite{bk1,chamon,nayak} rather than  metallic\cite{finkel}.  However, recent experiments\cite{metalreview} on 2D semiconductor heterostructures indicates
that in the strongly interacting regime,
a metallic state forms. At present, this problem is
 unresolved\cite{metalreview}.  The construction of bosonic or electronic metallic states in 2D 
remains one of the grand challenges in solid state physics.  We are concerned in this review with the possible existence of a metallic state for bosons in 2D.
 
\noindent {\bf Phase and Particle Number Duality}

To understand why a Bose metal poses a serious theoretical question,
we review the two standard ways in which superconductivity can be destroyed.
Pair formation and phase coherence underlie superconducting order.  Consequently, we represent the order parameter,  $\Psi(\vec r)=\Delta\exp(i\phi(\vec r))$, for a superconductor as a simple product of the pair amplitude, $\Delta$ and a rigid time-independent phase, $\phi(\vec r)$.  In terms of the original electron coordinates,
\beq                                    
\Psi(\vec r)=\langle c_{\downarrow}(\vec r)c_{\uparrow}(\vec r)\rangle=\langle\psi(\vec r)\rangle,                          
\eeq
where $c_{\sigma}(\vec r)$ annihilates an electron with spin $\sigma$ at $\vec r$, the angle brackets indicate an average over the quantum states of the system and $\psi(\vec r)$ is the wavefunction for a Cooper pair. Phase coherence 
occurs
when the phase correlator $G(\vec r)=\langle\psi(\vec r)\psi(0)\rangle$ approaches a non-zero value for large
 $\vec r$.  However, this rule is too restrictive.   In 2D, $G(\vec r)\propto r^{-\eta}$ decays algebraically\cite{bkt1,bkt2} at the finite temperature superconducting transition, with $0<\eta<1$. 
Herein lies a key difference between superconductivity in 2D and bulk systems.
  Because of the algebraic fall-off of phase coherence in 2D, the temperatures at which the Cooper pair amplitude
first becomes non-zero and global phase coherence occurs can differ substantially.  We will refer to the Cooper pair onset
temperature as $T_{c0}$ and the phase ordering temperature as $T_c$. 

\begin{figure}[t]
\centering
\includegraphics[width=5cm]{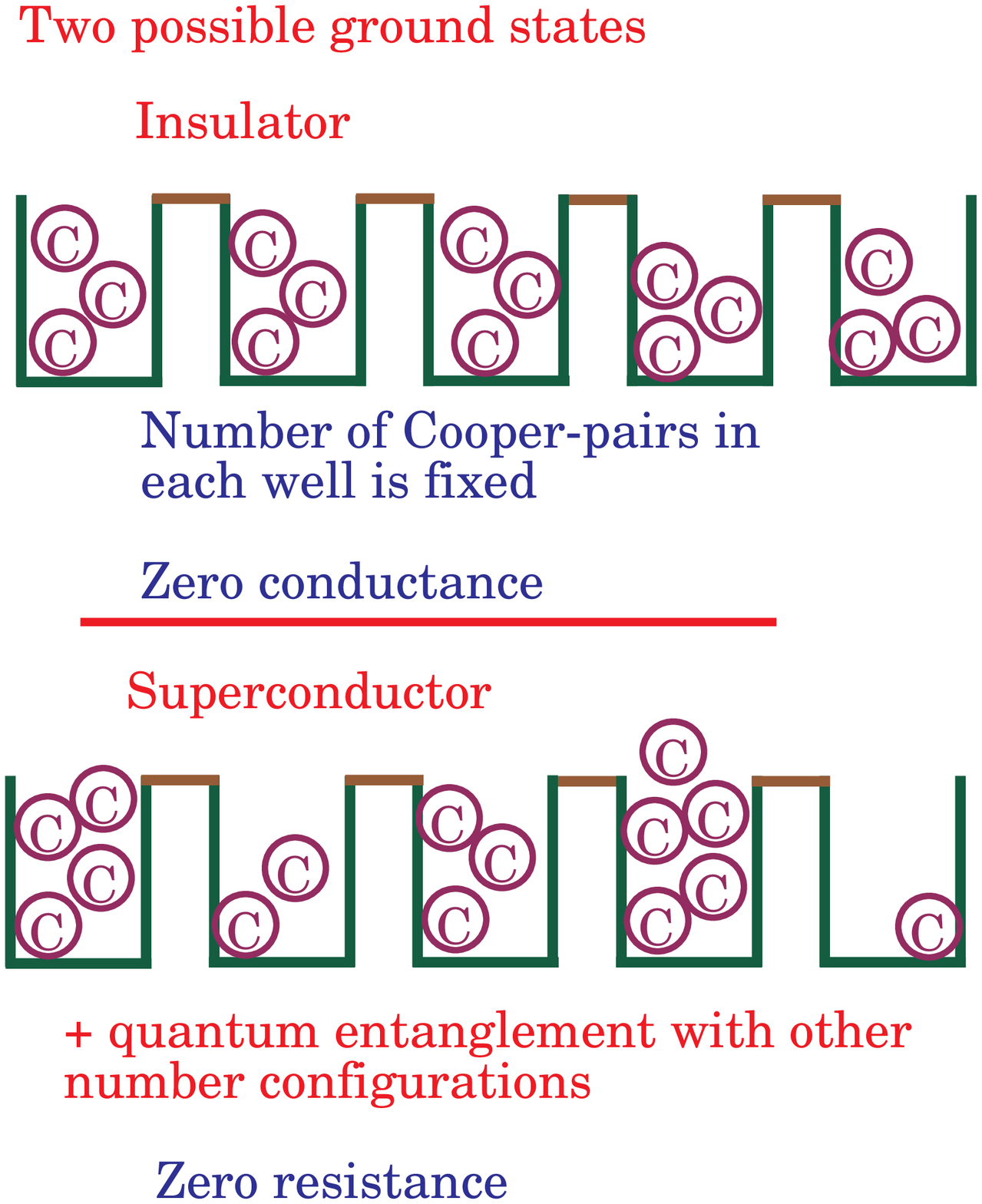}
\caption{Insulating and superconducting ground states of Cooper 
(C) pairs illustrating the conjugacy between phase and number 
fluctuations of the Cooper pairs. In the insulator, Cooper pair 
number fluctuations cease leading to infinite uncertainty in the 
phase. Contrastly, in a superconductor, phase coherence obtains, 
leading thereby to infinite uncertainty in the Cooper pair 
particle number.} 
\label{fig1}
\end{figure}

Consequently, superconductivity is destroyed by either breaking the Cooper pairs or disrupting phase coherence.  As a result of the localization principle in
 2D, an insulating state necessarily results in either case:  single electrons are localized by disorder and 
phase-incoherent Cooper pairs are limited spatially by the localization length. However, an
even more profound quantum principle underlies localization in the latter case. In the localized boson regime, snap shots at varying times of regions of the sample on the scale of the 
localization length reveal no fluctuation in the number of Cooper pairs.  However, in the superconducting state, analogous snap shots reveal that the Cooper pair number varies wildly (Fig. \ref{fig1}). The simple reason is that phase and particle number are conjugate variables and hence, their simultaneous
measurement is limited by the Heisenberg uncertainty principle.  Consequently,
bosons can either be in an eigenstate of particle number or phase.  The eigenstate of phase is a superconductor and that of particle number is a localized insulator.  No other possibilities exist.  At the transition point between the insulator and superconductor,
Cooper pairs on the brink of losing phase coherence diffuse\cite{matthew,herbut}
with a conductance whose scale is set by the quantum of conductance for charge 2e bosons,
namely $(2e)^2/h$ where $h$ is Planck's constant.  The coefficient of this 
conductance is not necessarily unity\cite{nayakchamon} as will be seen.

Experimentally, 
one transforms a superconductor into an
insulator by changing the intensity of the laser light
in optical lattices\cite{optical} or for thin films by either decreasing the film thickness\cite{jaeger,gl} or applying a perpendicular
magnetic field\cite{hebard,yazdani,ephron,mason2,mooij}. Decreasing the thickness increases boundary scattering and hence disorder
drives the transition, whereas in the application of a magnetic field,
 resistive topological excitations called vortices (the dual of 
Cooper pairs) frustrate the onset of global phase coherence. As long
as the two ground states on either side of the transition correspond to condensed and localized bosons, the transition
is controlled by quantum mechanics
 in the sense that 
a $T=0$ quantum critical point governs the physics.  Consequently,
the physics should be independent of the tuning parameter. 
Early experiments on the field-tuned insulator-superconductor
transition (IST) in $InO_x$\cite{hebard} films and disorder-tuning in
Bi deposited on Ge\cite{gl} seemed to fit the 
expected paradigm shown in Fig. (\ref{fig1}).  Namely, below a critical value
of the magnetic field or the disorder,  a superconducting state obtained with a vanishing resistance and beyond, the resistivity turned upwards as the temperature
was decreased indicating an insulating state, in apparent with theory. 

{\bf Experiment: Bose Metal}

\begin{figure}
\begin{center}
\includegraphics[width=6cm]{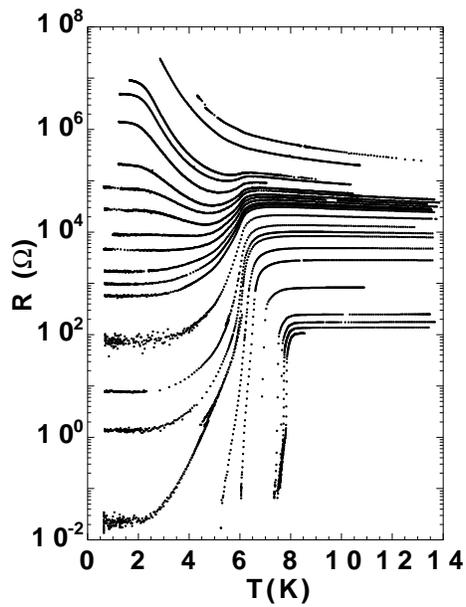}
\caption{Reprinted from C. Christiansen, L. M. Hernandez, and A. M. Goldman,
{\it Phys. Rev. Lett.}, {\bf 88}, 37004 (2002). Evolution of the temperature dependence of the resistance for
a series of Ga films. Film thicknesses range from 12.75 A to 16.67A and increases from top to bottom.  The leveling of the resistance once superconductivity
is destroyed (zero resistance curves) is not consistent with conventional
wisdom. Note that the plateau value of the resistivity increases 
as the distance from the superconducting phase increases. } 
\label{fig2}
\end{center}
\end{figure}

\begin{figure}[!t]
\begin{center}
\includegraphics[height=7cm]{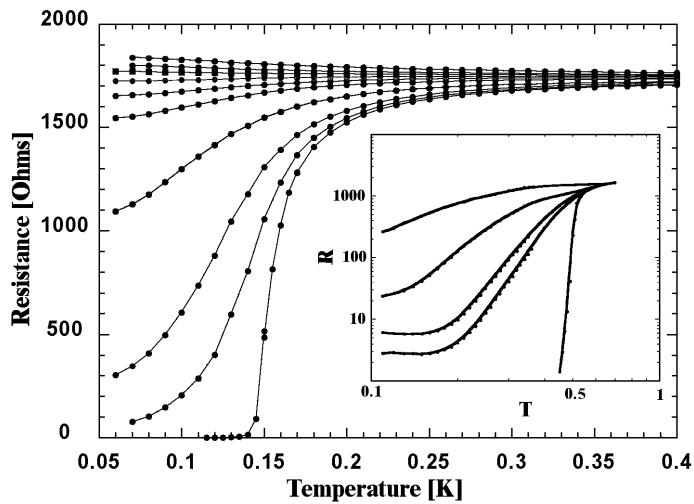}
\caption{Electrical resistance of MoGe thin film plotted vs 
temperature at B=0, 0.5, 1.0, 2.0, 3.0, 4.0, 4.4, 4.5, 5.5, 6 kG. 
The sample becomes a superconductor at 0.15 K in zero field but 
for fields larger than about 4.4 kG the sample becomes insulating. At fields 
lower than this but other than zero, the resistance saturates. The 
saturation behavior is better shown in the inset for another 
sample with a higher transition temperature. The inset shows data 
for B= 0, 1.5, 2, 4,and 7kOe. At higher field, this sample is an 
insulator. Main figure reprinted from A. Yazdani and A. Kapitulnik,  Phys. Rev. Lett. 
{\bf 74}, 3037 (1995), while the inset is from, Phys. Rev. Lett. {\bf 76}. 1529 (1996).} 
\label{ykap}
\end{center}
\end{figure}

\begin{figure}
\begin{center}
\includegraphics[height=8cm]{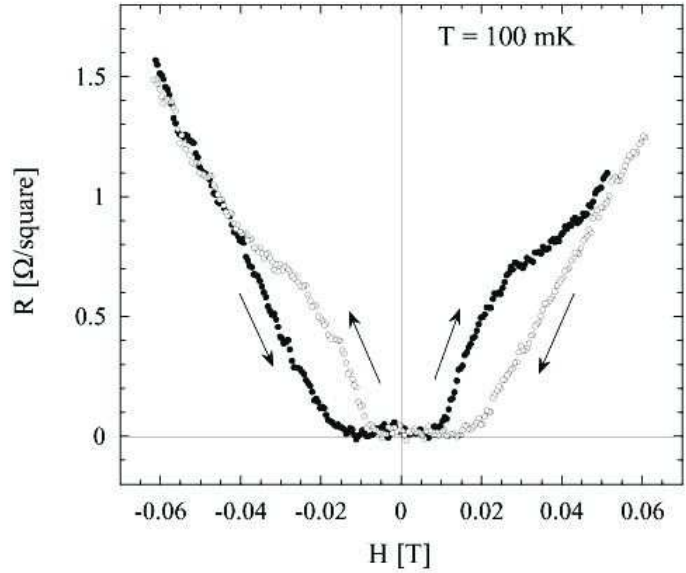}
\caption{ Reprinted from {\it Phys. Rev. B},
{\bf 64}, 60504-1 (2001).  Magnetoresistance at $T=80mK$ near the lower critical field, $H_M=0.018T$ of a
40\AA sample of Mo$_{43}$Ge$_{57}$ sandwiched between insulating layers
of amorphous Ge on SiN substrates. Arrows indicate the direction of the field
sweep.  The curves are all shifted by 87 Oe to account for trapped flux in 
the 16 T magnet.  The vanishing of the resistivity at $H_M=0.018T$ is interpreted as the transition to the true superconducting state. } 
\label{fig3}
\end{center}
\end{figure}

However, such agreement did not last long.  As early as 1989, Goldman and collaborators\cite{jaeger} observed that in a wide array of homogeneously
disordered films ranging from Ga, Al, Pb, to In, continuously decreasing
the thickness led first to a leveling of the resistance at low temperatures
immediately after superconductivity was destroyed.  Homogeneously disordered films, that is films in which the disorder is on atomic scales, can be produced by sputtering or from films deposited on specially prepared substrates cooled 
to liquid helium temperatures\cite{jaeger}.  
 The data shown
in Fig. (\ref{fig2}) taken on Ga deposited on Alumina demonstrate that the leveling extrapolates smoothly to $T=0$, indicating the emergence of a true metallic phase.  Unlike the earlier films which were homogeneous, the Ga/alumina
films were granular.  Let $g$ be the coupling constant (in this case the amount of disorder) that drives the destruction of superconductivity and $g_c$ the critical
value at the onset of the metallic state.  Fig. (\ref{fig2}) suggests that 
the $T=0$ value of the resistivity in the metallic phase is critical
in that it scales as some power of $(g-g_c)$, thereby vanishing at criticality.
This observation will
play a central role in constructing a theory for the metallic phase. For sufficiently 
thin films, insulating behavior obtains.   What about field-tuning?  In a magnetic field, vortices of only one vorticity are
possible in the superconducting phase.  At low temperatures and
non-zero magnetic fields, an exponentially
small resistivity of the form $R(T) \propto \exp{-C|\ln T|/T}$ is expected on 
the superconducting side as a result
of thermally assisted quantum tunneling of vortices.  However,
several groups\cite{yazdani,ephron,chervenak} have observed a leveling (see Fig. (\ref{ykap})) of the resistance here as well rather than 
the expected exponential dependence.  In fact, similar behavior has been
observed in graphite \cite{kopel}.
That the metallic state is really an artifact of failed refrigeration
has been addressed by minimizing power dissipation\cite{jaeger} and
by the observation that in the field-tuned samples\cite{ephron}, the effective temperature
associated with the high temperature activated behavior varied systematically
with magnetic field.  Further, the deviation\cite{ephron} from activated behavior
occurred at a lower temperature as the field increased, where power dissipation
and sample heating effects are greatest.  Hence, the metallic state is not 
a simple artifact of heating.
More recent 
experiments\cite{mason2} on the low-field regime reveal that true superconductivity
with a vanishing resistivity persists only until a magnetic field of magnitude $H_M=0.018T$.  The turn-on of the resistivity at $H_M=0.018T$ is shown
clearly in Fig. \ref{fig3}.  Insulating behavior does not obtain
until $H>H_c=1.8T$. Consequently, the metallic phase persists over a wide range
of magnetic field, $0.018<H<1.8T$.  Also apparent from Fig. \ref{fig3} is
 that the resistivity at a fixed temperature and magnetic field is not unique: 
cycling the magnetic field up and then down changes the
resistivity, an indication that the energy landscape has numerous metastable states as in the case of window glass, for example.

{\bf Electron Insulator}

Of course, the observation of a metallic phase poses an additional problem: 
What is the nature of the insulating phase?  This
problem surfaces because if a metallic state exists, then the insulator cannot arise simply by breaking phase coherence. Perhaps the pairs fall apart
to form the insulator?  This is precisely
what was\cite{valles} concluded from
tunneling measurements which yield a linear relationship with unit slope between the pair amplitude,
$\Delta$ and $H_c$.  This relationship holds only if $H_c$ corresponds to the field at which Cooper pairs disintegrate.  It was also noticed\cite{yazdani}
 that the value of $H_c$ was identical to the known value of the
magnetic field
at which Cooper pairs fall apart in MoGe.  Hence, it is reasonable to conclude\cite{fink,philmag} that it is electronic excitations rather than localized Cooper pairs that ultimately drive the insulating state at least in the homogeneously disordered films. On this reading, it is the metal that corresponds to the breaking of phase coherence and as a result, bosons carry the current.

{\bf Proposed Theories}

Hence, the problem at hand is the construction of a theory for a Bose metal
in 2D.  Such a metal must be stable to disorder and have
a resistivity that scales (see Fig. \ref{fig2}) as some power of the distance from
criticality, $g-g_c$.  Because the basic excitations are Cooper pairs, models with superconducting grains have been used widely to solve this problem\cite{spivak,spivak1,dd,wagen}.  In such models, Cooper pairs tunnel between
the grains with an amplitude, $J$, and the charging energy of each grain
is $E_C$.   Some have put aside
the issue of the origin of the metallic state\cite{spivak1,spivak} and
considered instead an idealized model of superconducting grains embedded in a
metal.  While the superconducting side is complicated
by electron-hole pair scattering from normal electrons\cite{spivak1,spivak}, quenching of superconductivity simply
unmasks the background conductance of the host metal.  Hence, by
design, such models admit a direct transition from a superconductor to a metal.
In the absence of the host metal, such hybrid models with normal electrons providing dissipation in an array of superconducting grains exhibit metallic behavior only at the separatrix between the insulator and the superconductor\cite{wagen}.  In other models\cite{dd} Coulomb interactions between the superconducting grains were included.  In this vein, it has been proposed\cite{dd} that beyond a critical grain size, a metallic phase obtains.  This model raises the 
possibility that short-range Coulomb interactions lead to new physics near
the quantum critical point associated with the destruction of phase coherence in an array of superconducting grains. Scaling arguments\cite{scal} indicate, however, that short-range interactions are irrelevant near the corresponding quantum critical
point as long as the phases are translationally and rotationally invariant as would be the case for a Bose metal. Additionally, the Bose superfluid has been shown
to form either a toroidal magnet or a Fermi liquid in the presence of strong
Coulomb repulsions, neither phase of which is metallic in 2D\cite{larkin}.  
The yet more exotic model of
bosons moving on a square lattice with ring exchange, that is, cyclical permutations of bosons on four mutually neighboring lattice sites has been considered\cite{ring}.  While the resultant phases\cite{ring} that are generated
share much in common with metals, such as gapless excitations, a finite compressibility, and the Bose analog of a Fermi surface, the conductivity vanishes as $T^\gamma$ where $\gamma>0$ and hence metallic conductivity does not survive at $T=0$.  Alternatively,
we have pointed out\cite{dd1} that the localized `insulator'
actually has a finite conductivity at zero temperature as a result
of a subtle cancellation between the exponentially small density
of states and the exponentially long mean-free path for bosonic quasiparticles. This result is irrelevant to the experiments
because any amount of disorder reinstates the insulator. 

{\bf Bose Metal: Phase Glass}

\begin{figure}
\begin{center}
\includegraphics[width=6cm]{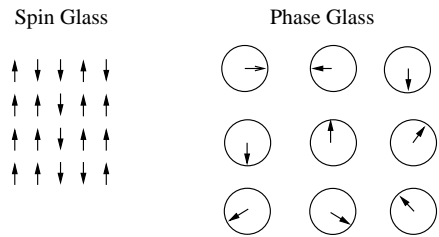}
\caption{ Patterns of spin and phase ordering in a spin glass and in a phase or rotor glass.  In a spin glass, the spins point up or down but randomly such that the net
magnetization vanishes.  In the rotor glass it is the phases that point along random non-collinear directions in the $x-y$ plane, for example. } 
\label{glass}
\end{center}
\end{figure}

So how can the experiments be explained?  Some have advocated that dissipation might be the key\cite{dissip,kpcoll}. However, it is difficult to pinpoint an external source of dissipation that survives
at zero temperature. Further, recent work\cite{dp4,nl} suggests that the standard treatment of dissipation in models for phase fluctuations in an ordered array of superconducting grains cannot produce a metallic phase at $T=0$ even
in the presence of electronic excitations\cite{wagen}.  Consequently, it is advisable to focus on new phases
in which dissipation is self-generated.  As an example, consider
 the quantum phase glass model recently proposed\cite{dp2} in which
disorder in the distribution of tunneling amplitudes, $J_{ij}$ and quantum fluctuations destroy phase coherence.  To describe superconductivity and to capture
 the frustration
required for glassy ordering, the distribution of $J_{ij}$'s must have a non-zero mean and contain both $\pm J$, respectively\cite{dp4}.  
While it might seem strange that 
the the sign of $J$ is random, this state of affairs is inextricable from
the role disorder plays in a superconductor as a direct consequence\cite{sk} of exchange
effects arising from the transport of a Cooper pair through a
localized defect. Hence, models with purely on-site disorder are inadequate.  While global phase coherence is absent
from the glassy phase, locally the phase on each lattice site points along
a fixed direction.  However, the directions differ from site to site as illustrated in Fig. \ref{glass}.  
 Consequently, in the phase or rotor glass,
 the phases on each site are frozen along mutually non-collinear directions such that $\langle \exp(i\phi(r_j))\rangle\ne 0$ but $\sum_j\langle \exp(i\phi(r_j))\rangle=0$. Here, the angle brackets indicate an average over the distinct
quantum states of the glass.  The effective order parameter\cite{by} for the glassy phase,
$Q(\tau)=\langle \exp(i\phi(r_j,0))\exp(i\phi(r_j,\tau)\rangle\ne 0$, reflects the local breaking of spin rotation invariance. Classically,
 $Q(\tau)$ relaxes exponentially fast to its equilibrium value.
However, quantum mechanically\cite{huse,sachdev}, $Q(\tau)$ decays
as $\tau^{-2}$. Such sluggish
phase relaxation gives rise to a density of low-lying excitations
that scales as $|\omega|$ and hence exceeds
the standard $\omega^2$ dependence of the density of states in
a superconductor.  Such low-lying excitations
have two key profound effects. First, the effective dimension
of the phase glass is now $d_{\rm eff}=2+2=4$, the additional dimensions arising from the time dynamics.  Second, in the total free energy in the glassy
phase, such excitations couple directly to the bosonic excitations from the fluctuations of the superconducting order parameter.
Bosons moving in such a glassy environment with a myriad of metastable states fail to localize\cite{dp2} because 
no true ground state exists.  The result\cite{dp2} is a metal at $T=0$ with
a critical resistivity that scales as $(g-g_c)^{p}$, where $p>0$, consistent with experiment (see Figs. \ref{fig2} and \ref{fig3}). This result holds even
when the bosons are allowed to interact\cite{dp2}.  Further, the vanishing of the resistivity at the critical
point is consistent with a recent analysis\cite{nayakchamon} that demonstrates that at criticality, the resistivity does depend universally on $h/4e^2$.  However, the prefactor
is zero\cite{nayakchamon}.  Hence, the phase glass is a candidate to 
explain the intervening metallic phase.

{\bf Is a Phase Glass Stiff?}

Is a phase glass a superconductor in disguise?  If yes, then the metallic behavior found
above, while intriguing, will be dwarfed by the infinite contribution to the conductivity arising from the collective mode (or phase stiffness) of the phase glass. This question
is relevant to topological glasses in general, in particular the vortex glass\cite{ffh} which has been argued to have zero resistance and to explain the ground state of copper oxide superconductors in a perpendicular magnetic field. 
The vortex glass state is currently controversial because extensive
experiments on copper oxide samples designed to optimize the conditions
for the validity of the vortex glass model, namely untwinned  (that is no grain boundaries and hence the defects are point-like) YBa$_2$Cu$_3$O$_{7-\delta}$ (YBCO) crystals, find an absence of scaling and a non-vanishing linear resistivity
below the putative glass transition\cite{ybco,scaling}.  
While vortex and phase glasses are not identical, the conclusions
regarding the stiffness should be model independent simply because global rotation of all the phases (or spins) around any particular axis (a generator of the group SO(m)) leaves the free energy invariant.  However, the rotated and unrotated states are distinct. As a consequence, glasses of this type break
SO(m) symmetry and a gapless mode should exist.  Technically, the phase stiffness is defined by applying a twist to all of the spins and determining the resultant change in the free energy.  If what results can be
 written simply as  
a ballistic mode dispersing as a linear function of momentum, as in the dispersion of light, for example, then a phase stiffness exists.  However,
all exact calculations at long-times in
either the Heisenberg\cite{ft,kotliar2} or quantum rotor glass\cite{dpstiff}
find that the equilibrium stiffness vanishes.  Nonetheless,
there is a massless mode\cite{ft,kotliar2,dp2} that disperses as $\omega \propto i k^2$.  At short times where the spin glass lives in a single valley,
the stiffness\cite{kotliar1} is non-zero.  It is this
single-valley view (infinitely high
barriers) that undergirds the vortex glass\cite{ffh}.  The physical reason for this difference is simple (see Ref. 44, p. 945) in that a spin glass is stiff if it is not allowed to hop (tunnel in the quantum rotor glass) from valley to valley as the spins are twisted.  Hence, in a full statistical mechanical calculation in which
all possible configurations are included as the twist is applied, the stiffness vanishes\cite{ft,kotliar2,dpstiff,by}. 
 So which result is relevant
to experiments?  In a true DC measurement, $\omega\ll\omega_g(T)$, where $\omega_g(T)$ is the barrier hopping frequency, 
the full statistical mechanical treatment
is relevant and no stiffness should be expected.  Although $\omega_g(T)$ is not
known, the systematic fluctuations in the leveled region of the conductivity in Fig. \ref{fig2} indicate that the measurements are in the true DC limit.
Similarly, in the mixed state
of the copper oxides, the non-vanishing of the 
resistivity below\cite{ybco,scaling,vm} the putative vortex glass transition suggests that it is the
equilibrium stiffness that is relevant. 

{\bf Final Remarks}

The observation of a metallic phase in thin metal alloy
films requires a new state of matter that does not fall prey
to the standard localization principles for electrons or bosons.  In the presence of disorder, superconductors lose phase coherence and become
glassy. Bosons moving in the resultant glassy background remain metallic
at $T=0$ provided that the low-lying excitation spectruum scales as a linear function of frequency as in the phase glass. The corresponding result for the vortex glass is currently not known.  Nonetheless, as we have seen, whether topological glasses have a phase stiffness
is a matter of time scales.  All such glasses (the vortex glass included) lose their single-valley stiffness in the long-time limit.  Perhaps
the persistence of a linear current-voltage response in YBCO below
the putative vortex glass transition\cite{ybco,scaling} is fundamentally
telling us that the configuration of the vortices in the glassy state is
not static and perhaps vortices freeze more like window glass as has been advocated\cite{vm}. Hence, our proposal that the Bose metal is glassy resonates with the experimental results on the mixed state of the cuprates.  While the hysteresis shown
in Fig. \ref{fig3} is symptomatic of glassiness,
additional experiments such as noise, relaxation time measurements, as well as $T\rightarrow 0$ transport studies on the mixed state of the cuprates
are needed to confirm the Bose metal picture.   Indeed, glasses with their intrinsic low-lying excitations offer a glimpse into the route by which bosons circumvent localization in the presence of disorder. 

We thank A. Goldman and D. S. Fisher for useful discussions and the NSF Division of Materials Research and the ACS Petroleum Research Fund for supporting this research and S. Sachdev for Fig. 1.

\newpage
\bibliography{scibib}

\bibliographystyle{Science}

\newpage
\noindent
\end{document}